\documentclass[11pt]{article}

\usepackage{amsmath}
\usepackage{amsthm}
\usepackage{amssymb,amsfonts}
\usepackage{array} 
\usepackage{float}

\usepackage[mathscr]{eucal}
\usepackage[english]{babel}
\usepackage[utf8x]{inputenc}
\usepackage[T1]{fontenc}
\usepackage{graphicx}
\usepackage{epstopdf}
\usepackage{geometry}
\usepackage{color}
\usepackage{times}
\usepackage{hyperref}
\usepackage[round, authoryear]{natbib}
\DeclareMathOperator*{\argmin}{argmin}
\DeclareMathOperator*{\argmax}{argmax}

\def \E{\mathbb{E}}
\def\alp{\alpha}
\def\eps{\epsilon}
\def\gam{\gamma}
\def\Gam{\Gamma}
\def\sig{\sigma}
\def\Sig{\Sigma}

\def\PhDperp{P_{\hat{D}}^{\perp}}
\def\lam{\lambda}
\def \R{\mathbb{R}}
\def \rU{\text{U}}
\def \Var{\text{Var}}
\def\bSig\mathbf{\Sigma}

\newtheorem{theorem}{Theorem}[section]
\newtheorem{lemma}[theorem]{Lemma}

\newtheorem{corollary}[theorem]{Corollary}
\newtheorem{assumption}[theorem]{Assumption}

\title{Mendelian Randomization when Many Instruments are Invalid: Hierarchical Empirical Bayes Estimation}
\author{Sai Li\footnote{Department of Statistics and Biostatistics, Rutgers University. Address: 110 Frelinghuysen Road, Piscataway, NJ 08854. Email: sl1022@scarletmail.rutgers.edu.}}
\begin{document}
\date{}
\maketitle

\begin{abstract}
Estimating the causal effect of an exposure on an outcome is an important task in many economical and biological studies. Mendelian randomization, in particular, uses genetic variants as instruments to estimate causal effects in epidemiological studies. However, conventional instrumental variable methods rely on some untestable assumptions, which may be violated in real problems.
In this paper, we adopt a Bayesian framework and build hierarchical models to incorporate invalid effects of instruments. We introduce an empirical Bayes estimator for which some of the instruments are invalid by utilizing a Gaussian mixture prior. Theoretical performance and algorithm implementations are provided and illustrated. The reliable performance of the proposed method is demonstrated in various simulation settings and on real datasets concerning the causal effects of HDL cholesterol and LDL cholesterol on type 2 diabetes.
\end{abstract}

%





%

\section{Introduction}
\label{s:intro}

Inferring the causality between exposures and outcomes is a crucial task in social science and epidemiology. Mendelian randomization (MR) uses genetic variants as instruments to measure the causal effect of a specific exposure on an outcome \citep{MR1, MR2}. As a counterpart to the randomized controlled trial (RCT), MR can address areas where an RCT would be impossible or unethical. With more and more available genome-wide association studies (GWAS), researchers are able to find genetic variants which are robustly associated with target exposures and infer the causality between exposures and outcomes via the variation of genetic variants. 

For instance, some recent studies raise an intriguing question whether there exists a causal relationship between low-density lipoprotein (LDL) cholesterol and type 2 diabetes. Statin therapy has been shown to reduce cardiovascular disease by lowering LDL \citep{LDL1}.  However, it is associated with a $9\%$ increased risk for incident diabetes in RCT studies \citep{LDL2}. On the other hand, another LDL lowering drug, Evolocumab, which uses a different bological pathway, has not been shown to have a significant effect on the incident diabetes in RCTs \citep{LDL3}. Thus, it is of interest to study whether the increased risk of diabetes is caused by lowering LDL or as opposed to medication-specific effects. This problem is analyzed in this paper as a case study with MR methods applied on summary data from GWAS.

There are many advantages of genetic variants serving as instruments. Firstly, in genetic associations, the direction of causation is always from the genetic polymorphism to the phenotype of interest, and not vice versa. Secondly, genetic variants are subject to relatively small measurement error or confoundness, as opposed to conventionally measured environmental exposures, which are often associated with a wide range of behavioral, social and physiological confounding factors. Thirdly, MR is more cost-effective compared with RCTs.

On the other hand, some concerns are raised about applying the MR methods, such as weak instruments, the confoundness of genotype, and canalization. Using multiple instruments can increase the power of genotype-exposure and genotype-outcome association, but may also introduce issues with linkage disequilibrium and pleiotropy \citep{MR4,MR5}. 

\begin{figure}[H]
\centering
\includegraphics[width = 3in]{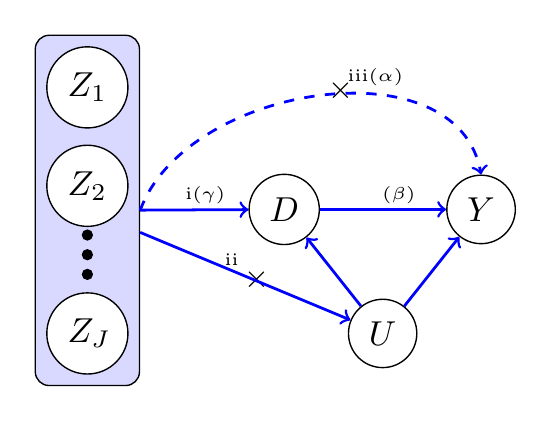}
\caption{Illustrative diagram of conventional instrumental variable assumptions and the relaxed assumptions in this paper. Crosses indicate violations of assumptions. Dashed arrows indicate the effects allowed to exist in this paper. Parameters in the parentheses correspond to the notations of such effects in model (\ref{eq1-1}).}
\label{fig1}
\end{figure}

In conventional instrumental variable literature, the classical assumptions for valid instruments (Figure \ref{fig1}) are \citep{MR6}
\begin{itemize}
\item[(i)] Relevance: The genetic variants $Z$ have an effect on exposure $D$.
\item[(ii)] Independence: The genetic variants $Z$ are uncorrelated with any confounders of the exposure-outcome relationship ($U$).
\item[(iii)] Exclusion restriction: The genetic variants $Z$ affects outcome $Y$ only through exposure $D$.
\end{itemize}

We say a genetic variant is valid if it satisfies assumptions (i)-(iii). However, not all of them are fully realistic. 

Assumption (i) can be fulfilled by selecting significant genetic variants from available GWAS, while (ii) and (iii) are both untestable and may be violated in MR. For example, (iii) is known to be problematic due to the pleiotropic effects of genetic variants, which means that one gene can influence two or more seemingly unrelated phenotypic traits. Additionally, genetic variants may have direct effects on the outcome. 

The purpose of this paper is to develop a reliable estimator of the causal effect free of assumption (iii). That is, corresponding to Figure~\ref{fig1}, we allow the effects represented by the dashed arrow to exist. 

Many recent works study the relaxation of assumption (iii) from various perspectives. \citet{Egger} introduced Egger's regression method under the InSIDE assumption (instrument strength independent of direct effect). \citet{Lasso} develop a Lasso-type estimator under some regularity conditions and the ``partially invalid" assumption, which means some genetic variants are valid.  \citet{IVWM} borrow tools from meta-analysis and develop a consistent estimator when at least $50\%$ of genetic variants are valid.  A pleiotropy-robust MR method is introduced by \citet{MR6} using a subsample which is independent of the exposure to estimate the pleiotropic effects. However, these types of assumptions can be hard to check in reality and hence restrict the applicability of such estimators.

In a closely related article, \citet{FG15} consider a hierarchical model to account for the randomness in data collection, unmeasured covariates, and treatment effect variation. However, their approach does not incorporate instrumental variables, while an MR problem is intrinsically equipped with genetic variants as instruments. 

There are other recent advances on extending the applications of MR, such as estimation with two-sample summary data \citep{Twosample1}, the study of power and instrument strength requirements \citep{Power}, and pathway identification \citep{Network}. 

In this paper, we use the empirical Bayes hierarchical models to incorporate the pleiotropic effects. Some noteworthy features of the proposed approach are:
\begin{itemize}
\item[(i)] The proposed method does not rely on the ``partially invalid" assumption or the InSIDE assumption, which are required in the many existing literatures and may not be biological plausible.
\item[(ii)] The proposed method is reliable even when there exist unbalanced pleiotropic effects and partial invalidness.
\item[(iii)] The estimation procedure is based on an easily implemented and computationally efficient algorithm, the Monte Carlo expectation-maximization (MCEM) algorithm.
\end{itemize}

The rest of this paper is organized as follows. In Section 2, we introduce notation, set up the model, and motivate our estimators. In Section 3, we establish the theoretical guarantees of our estimators and illustrate the implementations of the estimation procedure. In Section 4, we show how to apply our method with summary statistics. In Section 5, we apply our methods to both simulated experiments and real studies. Finally, we conclude the paper with some remarks and future research directions.

\section{Model set-up}
\subsection{Notation}
Observed genotypes are usually coded as the number of minor alleles, 0, 1, or 2.  Without loss of generality, we consider the case where the instruments are continuous in this paper. Let $Z_i\in \R^J$ denote the $i$-th observation  of $J$ instruments. Let $Z \in \R^{n\times J}$ be an $n\times J$ matrix of genetic variants whose $i$-th row consists of $Z_i$. Let $Z_j\in \R^n, j=1,\dots, J$ denotes the $j$-th column of $Z$. Let $D=(D_1,\dots, D_n)^T\in \R^n$, where $D_i\in \R$ is the $i$-th observation of the exposure. Let $Y=(Y_1,\dots, Y_n) \in \R^n$, where $Y_i\in \R$ is the $i$-th observation of the outcome. Note that we focus on quantitative type of exposures and outcomes in this paper. We assume that $Y$, $D$ and each column of $Z$ are centered for the analysis.

For a vector $r\in \R^d$, let $\|r\|_1= \sum_{i=1}^d|r_i|$ and $\|r\|_2= \sqrt{\sum_{i=1}^d r^2_i}$. For a matrix $G \in \R^{m\times d}$, let $P_G$ be the $m \times m$ orthonormal projection matrix onto the column space of $G$, i.e. $P_G = G(G^TG)^{-1}G^{T}$. Let $P^{\perp}_G = I_{m\times m}- P_G$. For a square matrix $G'\in \R^{d\times d}$, let $\Lambda_{\max}(G')$ be the largest eigenvalue of $G'$ and $\Lambda_{\min}(G')$ be the smallest eigenvalue of $G'$. Let $G'' \in \R^{d\times d}$ be another square matrix. We say $G' \preceq G''$ iff $G'' - G'$ is a positive definite matrix.

For a random variable $V\in \R$, let $\Var(V)$ denote the variance of $V$ such that $\Var(V) = \E[(V-\E[V])^2]$. Let $N(\mu,\sig^2)$ denote the Gaussian distribution with mean $\mu$ and variance $\sig^2$. Let $\Psi(\cdot)$ and $\phi(\cdot)$ be the cdf and pdf of a standard Gaussian random variable, respectively. Let $\rU[a,b] $ denote the uniform distribution on $[a,b]$ for $a<b$.

\subsection{Model specification}
Suppose that we observe i.i.d. copies of $(Z_i, D_i, Y_i), i=1,\dots, n$. 
We adopt the Neyman-Rubin's potential outcome framework \citep{NR2,NR1} and set up the model for observed data under assumptions (i) and (ii). 

With some basic derivations of the potential outcome model, we consider the following the model for observed data. For $i = 1,\dots,n$, 
\begin{equation}
\begin{cases}
D_i&=Z_i\gam+v_{i}\label{eq1-1}\\
Y_i&=\beta D_i+Z_i\alp+\eps_i, 
\end{cases}
\end{equation}
where $(v_i,\eps_i)$ has mean zero and covariance matrix $\begin{pmatrix}
    \sig^2_v & \sig^2_{v\eps}\\
    \sig^2_{v\eps} & \sig^2_{\eps}
    \end{pmatrix}$ conditioning on $Z$, $\beta \in \R$ is the causal effect of interest, $\gam = (\gam_1,\dots, \gam_J)^T\in \R^J$ with $\gam_j$ the strength of the $j$-th instrumental variable, and $\alpha = (\alpha_1,\dots, \alpha_J)^T\in \R^J$ with $\alpha_j$ the total effect of $j$-th genetic variant $Z_j$ on the outcome $Y$ not via the exposure $D$ or the common confounders $U$. For simplicity, we refer to $\alpha_j$ as the pleiotropic effect of $Z_j$. Moreover, the effect of the common confounders $U$ in Figure \ref{fig1} enters the model via $\sig^2_{v\eps}$.

When $\sig^2_{v\eps}$ is non-zero, $D$ is correlated with the error term $\eps$ in the exposure-outcome model. Therefore, the second equation in model (\ref{eq1-1}) does not satisfy the classical linear model assumptions. This issue can be taken care of by a well-established instrumental variable method, the Two-Stage Least Square (TSLS) estimation, provided that all of the genetic variants are valid, i.e. $\alpha = 0$. 

Specifically, one can construct a proxy of $D$, namely the least square estimate $\hat{D}$, such that
\begin{align}
\label{eq1-2}
   \hat{D} =  Z\hat{\gam},
\end{align}
where $\hat{\gam} = (Z^TZ)^{-1}Z^TD$. 

To ease the notation, let $\hat{v}_i = D_i-\hat{D}_i$ and $ \hat{\eta}_i = \eps_i + \beta \hat{v}_i$. We can rewrite the exposure-outcome model as
\begin{equation}
\label{eq1-3}
   Y_i = \beta\hat{D}_i + Z_i\alpha + \hat{\eta}_i.
\end{equation}
Note that $\E[Z_i^T\hat{\eta}_i]=0$ by the our assumption and construction. Thus, if $\alpha = 0$, model (\ref{eq1-3}) satisfies the moment condition 
$\E[Z^T(Y-\beta\hat{D})] = 0$, which sheds light on the TSLS estimator. Formally, define the TSLS estimator of $\beta$ as 
\[
\hat{\beta}^{(tsls)} = \argmin_{\beta\in \R} \|Y - \hat{D}\beta\|_2^2.
\]
It is easy to see that $\hat{\beta}^{(tsls)}$ is an asymptotically unbiased estimator of $\beta$ assuming that $\alpha = 0$ and $\|\hat{D}\|^2_2/n\rightarrow K_1>0$ as $n\rightarrow \infty$.

With $\alpha$ unknown and possibly nonzero, one may consider the multivariate least square estimator, say $(\check{\beta}^{(tsls)},\check{\alpha}^{(tsls)})$, such that
\begin{equation}
\label{eq1-5}
(\check{\beta}^{(tsls)},\check{\alpha}^{(tsls)}) = \argmin_{(\beta,\alpha)\in \R^{J+1}} \|Y - \hat{D}\beta - Z\alpha\|_2^2.
\end{equation}
However, it can be seen from (\ref{eq1-2}) that $\hat{D}$ is a linear combination of $Z$ and hence the column space of matrix $(\hat{D},Z)\in \R^{n\times (J+1)}$ is rank deficient. As a result, the parameter of interest, $\beta$, cannot be not identified.

In this paper, we consider a variation of (\ref{eq1-5}), which can be formulated as the regularized regression approach, i.e.
\begin{equation}
\label{eq1-6}
(\tilde{\beta}^{(2sls)},\tilde{\alpha}^{(2sls)}) = \argmin_{(\beta,\alpha)\in \R^{J+1}} \|Y - \hat{D}\beta - Z\alpha\|_2^2 + R_{\lam}(\alpha),
\end{equation}
where $R_{\lam}(a)$ is a regularization term indexed with $\lam$. 

The regularized estimators resulted from (\ref{eq1-6}) are equivalent to the posterior mode under a noninformative prior of $\beta$ and $\pi(\alpha|\lam)$ of $\alpha$, where $R_{\lam}(a) = -\log \pi(\alpha|\lam)$ \citep{SS3}. In addition, the square loss function in (\ref{eq1-6}) corresponds to the negative logarithm of the standard Gaussian density for $\hat{\eta}_i$.

Inspired by this equivalence and in order to adapt to general assumptions on $\alpha$, we adopt the Bayesian framework and manipulate the regularization term by specifying some flexible hierarchical priors on $\alpha$.

\section{Handling nuisance parameters: hierarchical models}

The hierarchical model is an effective tool for pooling information and simultaneous inference. A large class of shrinkage estimators are generated within this framework, such as the James-Stein estimator \citep{JS1} and the SURE estimator \citep{SURE1}. The risk properties of shrinkage estimators have been well-studied through a series of papers \citep{Baranchik70,Brown71,Str91}. 
\subsection{Gaussian prior with a data-driven location}
\label{sec: SG}
Effect sizes in genetics are often modeled under a Gaussian prior \citep{Gaussian}. When specifying the location parameter, we need to take into consideration the unbalanced pleiotropic effects, where the mean of the $\alpha$ is not zero. We specify the prior distribution of $\alpha$ as
\begin{equation}
\label{prior1}
    \alpha|\mu_{\alpha},\tau^2 \sim N(\mu_{\alpha},\tau^2I_{J\times J}),
\end{equation}
where $\mu_{\alpha}$ is unknown and $\tau^2$ is assumed to be known for the purpose of illustration.  

\subsubsection{Theoretical guarantees}  
We treat $\hat{\eta}_i$ as a Gaussian random variable with mean zero and variance $\sig^2_{\eta}$ to keep the form of square loss in the target function of (\ref{eq1-6}). Assume that $\tau^2$ and $\sig^2_{\eta}$ are known for the purpose of illustration. 



For some given $\mu_{\alpha}$, $\tau^2$, and $\sig^2_{\eta}$, let $ (\hat{\beta}^{\mu_{\alpha}},\hat{\alpha}^{\mu_{\alpha}})$ be the posterior mode under the prior (\ref{prior1}). We can obtain that
 \begin{equation}
 \label{func1}
(\hat{\beta}^{\mu_{\alpha}},\hat{\alpha}^{\mu_{\alpha}})= \argmin_{(\beta,\alpha)\in \R^{(J+1)}}\|Y-\hat{D}\beta-Z\alpha\|_2^2 + \frac{\sig^2_{\eta}}{\tau^2}\|\alpha-\mu_{\alpha}\|_2^2.
 \end{equation}
 Define two matrices $A$ and $B$, such that 
 \[
 A = \frac{1}{n\sig^2_{\eta}}Z^TP_{\hat{D}}Z~~ \text{and}~~ B =\frac{1}{n}(\frac{Z^TZ}{\sig^2_{\eta}} + \tau^{-2}I_{J\times J}).
 \]
\begin{assumption}
\label{ass1}
 Let $c^*$ be the largest eigenvalue of $AB^{-1}$ satisfying $0< c^* < 1$.
\end{assumption}
Next theorem provides an empirical error bound for $\hat{\beta}^{\mu_{\alpha}}$ defined in (\ref{func1}).
\begin{theorem}
\label{thm1}
Suppose that Assumption \ref{ass1} holds. For some given $\mu_{\alpha}$, $\tau^2$, and $\sig^2_{\eta}$, the absolute error of $\hat{\beta}^{\mu_{\alpha}}$ defined in (\ref{func1}) satisfies
 \begin{equation}
 \label{eq-thm1}
|\hat{\beta}^{\mu_{\alpha}}-\beta| \leq \frac{c^*\sig^2_{\eta}\|\hat{\gam}\|_2\|\alpha-\mu_{\alpha}\|_2}{\tau^2(1-c^*)\hat{D}^T\hat{D}}+\frac{c^*\|\hat{\gam}\|_2\|Z^T\PhDperp\hat{\eta}\|_2}{(1-c^*)\hat{D}^T\hat{D}}  +\frac{ |\hat{D}^T\hat{\eta}|}{\hat{D}^T\hat{D}}.
\end{equation}
\end{theorem}

Now we provide the conditions for $c^*$ to fall in the $(0,1)$ interval.
\begin{lemma}
\label{lem1}
If $0<\tau^2<\infty$, $0< \sig^2_{\eta}<\infty$, $\|\hat{D}\|_2>0$,  and $\Lambda_{\min}(Z^TZ/n) > 0$, Assumption 3.1 is satisfied.
\end{lemma}
It is not hard to see that for finite $\alpha_j -\mu_{\alpha}$ and $\hat{\gam}_j,~j=1, \dots, J$,  the right-hand side of (\ref{eq-thm1}) is of order $J/n$ under the assumptions of Lemma \ref{lem1}. Thus, $\hat{\beta}^{\mu_{\alpha}}$ is asymptotically unbiased for $J\ll n$ under our assumptions. 

Another important constant appears in the error bound is $c^*$, which can be viewed as a regularity constant on $D$ and $Z$. It can be seen that the error bound could become very large when $c^*$ is very close to 1. Thus, it is necessary to evaluate $c^*$, which in fact can be easily calculated given $\tau^2$, $\sig^2_{\eta}$, and the observations. It is simpler and calculable compared to the restricted isometry property (RIP) constants, which are used to bound the Lasso-type estimator in \citep{Lasso}. Moreover, this estimator does not depend on the number of valid instruments or the InSIDE assumption, which are always untestable in real applications.

Moreover, the performance of $\hat{\beta}^{\mu_{\alpha}}$ can be understood from the regularization perspective. From the formulation of (\ref{func1}) one can see that $(\hat{\beta}^{\mu_{\alpha}}, \hat{\alpha}^{\mu_{\alpha}})$ is a ridge-type regression estimator with penalty factor $\sig^2_{\eta}/\tau^2$ and a drift term $\mu_{\alpha}$. It is well known that ridge-type penalty has variance stabilization effect in the scenario of collinearity and shrinkage effect towards 0. By adding a ridge-type regularization term, we manipulate the variance-bias trade-off and achieve an asymptotically unbiased estimator under mild conditions. One should also notice that only the nuisance parameters $\alpha$ are penalized, not the parameter of interest $\beta$, which excludes the bias directly caused by regularization. Finally, the drift term $\mu_{\alpha}$ as well as the penalty term $\sig^2_{\eta}/\tau^2$ are always unknown and need to be estimated.

Now we formally set up the hierarchical structure for the observed data and involved parameters.

\subsubsection{The empirical Bayes hierarchical model}
For the unobserved variance components $\tau^2$ and $\sig^2_{\eta}$, a common choice is to place inverse-gamma priors on them. As one can see from (\ref{func1}), $\tau^2$ and $\sig^2_{\eta}$ together play a role in tuning the penalty parameter $\sig^2_{\eta}/\tau^2$ and hence have an effect on the accuracy of the estimation but do not change the form of the estimator. Thus, the result of Theorem \ref{thm1} can still be applied with estimated $\sig^2_{\eta}$ and $\tau^2$.

\begin{figure}[H]
\centering
\includegraphics[width =2.5in]{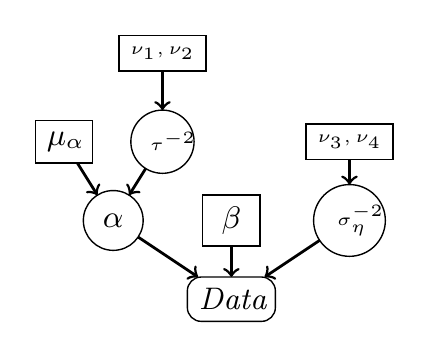}
\caption{The hierarchical structure of parameters for the hierarchical model (\ref{bayes1a})-(\ref{bayes1d}). All ``squares" represent parameters in the marginal distribution and all ``circles" represent represent parameters which only appear in the ``complete-data" posterior.}
\label{fig2}
\end{figure}

As illustrated in Figure \ref{fig2}, the fully Bayesian specification of the model is 
\begin{align}
Y|\hat{D}, Z, \beta,\alpha, \sigma^2_{\eta} &\sim N(\hat{D}\beta + Z\alpha, \sigma^2_{\eta} I_{n\times n}) \label{bayes1a}\\
 \alpha|\mu_{\alpha},\tau^2 &\sim N(\mu_{\alpha}, \tau^2I_{J\times J})\label{bayes1b}\\
\tau^{-2}|\nu_1,\nu_2&\sim \text{Gamma}(\nu_1,\nu_2)\label{bayes1c}\\
\sigma^{-2}_{\eta}|\nu_3,\nu_4 &\sim   \text{Gamma}(\nu_3,\nu_4), \label{bayes1d}
\end{align}
where $\beta$ and $\mu_{\alpha}$ are unknown, $\text{Gamma}(a,b)$ is Gamma distribution with shape parameter $a$ and rate parameter $b$, and $\nu_1-\nu_4$ are some prespecified constants based on our prior knowledge. In order to make the priors noninfluential, we set $\nu_1-\nu_4$ to be small constants during the estimation.
This hierarchical structure gives the marginal prior of $\alpha$ as
\[
    \alpha_j|\mu_\alpha,\nu_1,\nu_2\sim_{iid}T_{2\nu_1}(\mu_{\alpha}, \nu_2/\nu_1), ~j=1,\dots, J,
\]
where $T_{k}(\mu,\sig^2)$ is a Student's t-distribution with mean $\mu$, variance $\sig^2$, and degree of freedom $2\nu_1$.

For parameters $\beta$ and $\mu_{\alpha}$, we adopt the empirical Bayes framework and estimate them with the marginal maximum likelihood estimators. The empirical Bayes approach can be viewed as an approximation of a fully hierarchical Bayes analysis \citep{CG90,CG91}. It allows a second-level model learning and has been widely used for combining information and multi-task learning in statistics and machine learning \citep{EB1,EB2, EB3}. This empirical Bayes set-up allows the estimates of $\alpha$ to share information and borrow strength from each coordinate \citep{EB4}.

\subsubsection{The MCEM algorithm}
We briefly explain the algorithm implementations in this section. Details can be found in the Appendix. We adopt a variation of expectation-maximization (EM) algorithm, the Monte Carlo EM (MCEM) algorithm, to achieve the estimation \citep{,MCEM1, MCEM2,MCEM3}. The EM algorithm is a widely-adopted and computationally fast algorithm for handling missing values and unobserved variables. When the expectation in the ``E-step" does not have a closed form, one can approximate it with Monte Carlo samples. The MCEM algorithm iteratively estimates the marginal parameters and samples the middle-layered parameters until it converges.  (See  \citep{MCEM4} for a general description.)

Let $\mathcal{D}$ denote the observed data $(Z_i,D_i,Y_i)$ for $i=1,\dots,n$. The optimization starts with some initial value $ (\hat{\beta}^{(0)},\hat{\mu}^{(0)}_{\alpha})$. At $t$-th iteration $(t=1,2,\dots)$,
 generate $(\alpha^{(t)}_i ,(\tau^2)^{(t)}_i, (\sig^2_{\eta})^{(t)}_i)$, $i=1,\dots, m$, from the posterior distribution of $(\alpha,\tau^2,\sig^2_{\eta})$ under the current estimate $(\hat{\beta}^{(t-1)},\hat{\mu}^{(t-1)}_{\alpha})$. Then calculate the Monte Carlo estimate of $\E_{\alpha, \tau^2,\sig^2_{\eta}}[\log p(\beta,\mu_{\alpha}|\mathcal{D},\alpha, \tau^2,\sig^2_{\eta})]$, which is
\begin{align}
   \frac{1}{m}\sum_{i=1}^m \log p(\beta, \mu_{\alpha}|\mathcal{D},\alpha^{(t)}_i, (\tau^2)^{(t)}_i,(\sig^2_{\eta})_i^{(t)}).\label{eq3-3}
\end{align}

At the M-step, we compute the maximizer of (\ref{eq3-3}), i.e.
\begin{equation*}
   (\hat{\beta}^{(t)},\hat{\mu}^{(t)}_{\alpha}) = \argmax_{(\beta,\mu_{\alpha})\in \R^2}  \frac{1}{m}\sum_{i=1}^m \log p(\beta,\mu_{\alpha}|\mathcal{D},\alpha^{(t)}_i, (\tau^2)^{(t)}_i,(\sig^2_{\eta})_i^{(t)}).
\end{equation*}
At the convergence of $(\hat{\beta}^{(t)},\hat{\mu}^{(t)}_{\alpha})$, produce $\hat{\beta}^{(t)}$ as the final estimate of $\beta$. 
\subsection{A mixture Gaussian prior}
\label{sec: MREB}
If, in fact, some genetic variants are valid instruments, i.e. $\alpha_j =0$ for some $j \in \{1,\dots, J\}$, a single Gaussian prior may not be the appropriate structure to put on $\alpha$. A prior distribution which can induce a sparse posterior mode is preferable.

In the Bayesian framework, the ``Spike-and-Slab" prior \citep{SS1} is a well-established Bayesian variable selection procedure via a sequence of papers \citep{SS1,SS2,SS4,SS3}. It consists of a spike component and a slab component both centered at 0. 
This prior imposes larger shrinkage effects on the relatively small estimates and smaller shrinkage effects on the relatively large estimates. 

Our goal here is estimation rather than variable selection and our strategy is to add a spike component to a Gaussian component with an unknown center.
More specifically, we consider a mixture Gaussian prior as
\begin{align}
   \alpha_j|\mu_{\alpha},\xi_j,\tau^2,\nu_0&\sim_{ind} N(\xi_j\mu_{\alpha}, \nu_0\tau^2 + (1- \nu_0)\xi_j\tau^2)\label{prior2a}\\
   \xi_j|p_0 &\sim_{iid} \text{Ber}(p_0)\label{prior2b},
\end{align}
where  $\mu_{\alpha}$ and $\tau^2$ are unknown parameters, $\nu_0$ is a very small constant, and $\text{Ber}(p_0)$ is a Bernoulli distribution generating 1 with probability $p_0$.

If $\xi_j=1$, the prior distribution of $\alpha_j$ is $N(\mu_{\alpha},\tau^2)$; if $\xi_j = 0$, the prior distribution of $\alpha_j$ is $N(0,\nu_0\tau^2)$ with a small constant $\nu_0$ (say 0.001 as in the simulation). The parameter $p_0$ can be interpreted as the overall level of sparsity. 

\begin{figure}[H]
\centering
\includegraphics[width = 5in, height = 1.4in]{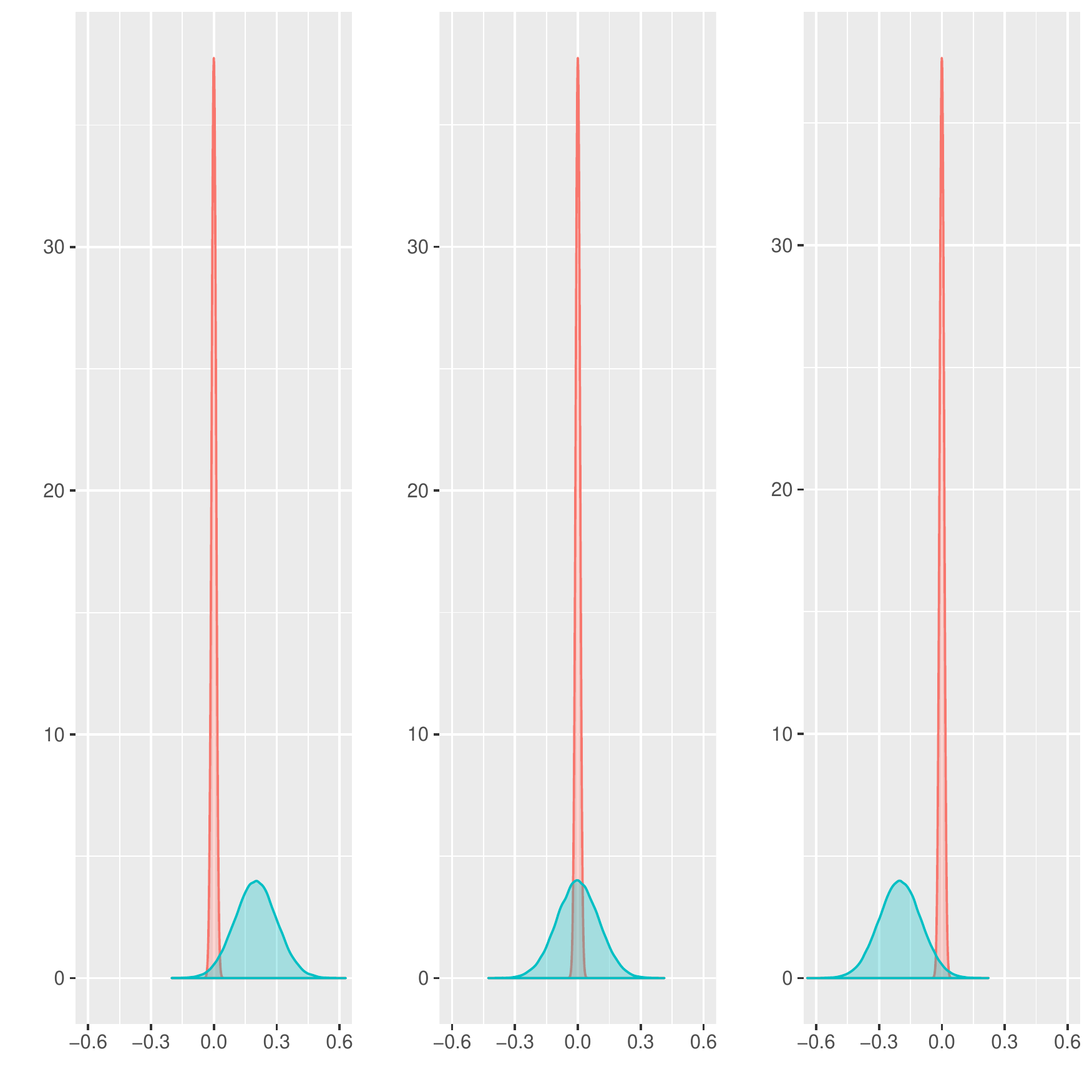}
\caption{Density plot of 10000 realizations of $\alpha_j$ from the mixture distribution (\ref{prior2a})-(\ref{prior2b}) with $p_0 = 0.8$, $\tau^2 = 0.01$, $\nu_0 = 0.001$, $\mu_{\alpha} = -0.2, 0$, and $0.2$ from left to right.}
\label{fig3}
\end{figure}

The priors in (\ref{prior2a}) and (\ref{prior2b}) allow a data-driven location of the ``slab" component as well as a data-driven sparseness parameter. In the same spirit as a single Gaussian prior, the purpose of fitting the mean of the nonzero component is to deal with unbalanced pleiotropic effects and reduce the estimation error. The purpose of a data-dependent $p_0$ is to allow the determination of the unknown sparsity. 
This formulation is able to incorporate the sparse structure of $\alpha$ and deal with the unbalanced pleiotropic effects at the same time (see Figure \ref{fig3} for an example).

Given $\mu_{\alpha}$, $\xi$, $\tau^2$, and $\sig^2_{\eta}$, the posterior mode under prior (\ref{prior2a}) can be written as
\begin{align}
(\hat{\beta}^{(\mu_{\alpha},\xi)}, \hat{\alpha}^{(\mu_{\alpha},\xi)}) = \argmin_{(\beta,\alpha)\in \R^{J+1}}\|Y - \hat{D}\beta - Z\alpha\|_2^2 + \frac{\sig^2_{\eta}}{\tau^2}\sum_{j=1}^J \frac{(\alpha_j-\mu_{\alpha}\xi_j)^2}{\nu_0+(1-\nu_0)\xi_j}. \label{func2}
\end{align}
From the regularization perspective, (\ref{func2}) implies that each $\alpha_j$ with $\xi_j=0$ is shrunk towards 0 with penalty level $\sig^2_{\eta}/(\nu_0\tau^2)$ and each $\alpha_j$ with $\xi_j=1$ is shrunk towards $\mu_{\alpha}$ with penalty level $\sig^2_{\eta}/\tau^2$. 

Let $\Gam_{\xi}$ be a diagonal matrix with $(\Gam_{\xi})_{j,j} = (\nu_0+(1-\nu_0)\xi_j)\tau^2$ for $j=1,\dots J$ and $B_{\xi} = (Z^TZ/\sig^2_{\eta} + \Gam_{\xi}^{-1})/n$. We require the following assumption to hold.
\begin{assumption}
\label{ass2}
Let $c^{**}$ be the largest eigenvalue of $AB^{-1}_{\xi}$ such that $0< c^{**}<1$.
\end{assumption}
\begin{corollary}
\label{cor1}
Suppose that Assumption \ref{ass2} holds. For given $\mu_{\alpha}$, ${\xi}$, $\tau^2$, and $\sig^2_{\eta}$, the estimation error of $\hat{\beta}^{(\mu_{\alpha},\xi)}$ defined in (\ref{func2}) satisfies
 \begin{align}
|\hat{\beta}^{(\mu_{\alpha},\xi)}-\beta| \leq &\frac{c^{**}\sig^2_{\eta}\|\hat{\gam}\|_2\|\Gam^{-1}_{\xi}(\alpha-\mu_{\alpha}\xi)\|_2}{(1-c^{**})\hat{D}^T\hat{D}}+\frac{c^{**}\|\hat{\gam}\|_2\|Z^T\PhDperp\hat{\eta}\|_2}{(1-c^{**})\hat{D}^T\hat{D}}  + \frac{|\hat{D}^T\hat{\eta}|}{\hat{D}^T\hat{D}}. \label{eq-thm2}
\end{align}
\end{corollary}

Now we compare the error bound of mixture Gaussian prior in (\ref{eq-thm2}) with that of single Gaussian prior in (\ref{eq-thm1}). From the definition of $\Gam_{\xi}$, one can see that see that $\Gam_{\xi} \preceq \Gam$ with the same $\tau^2$. By some simple linear algebra derivation, one can show that $c^{**} \leq c^*$, which is preferred. On the other hand, if valid instruments are all correctly selected, that is, if $\xi_j = 0$ for $\alpha_j = 0~(j=1,\dots, J)$,  
\[
 |\Gam^{-1}_{\xi}(\alpha-\mu_{\alpha}\xi)\|^2_2 = \sum_{j:\xi_j = 1}\frac{1}{\tau^2}(\alpha-\mu_{\alpha})^2 \leq \|\Gam^{-1}(\alpha-\mu_{\alpha})\|^2_2 .
 \]
 Thus, the mixture prior can possibly improve the estimation accuracy for sparse $\alpha$, but also depends on the estimation of $\xi$ and $\mu_{\alpha}$.
 
For the estimation side, we again use (\ref{bayes1c}) and (\ref{bayes1d}) as priors for unknown variance components $\tau^2$ and $\sig^2_{\eta}$ and build a fully Bayesian hierarchical model together with (\ref{prior2a}) and (\ref{prior2b}). We call such estimator of causal effect the \textbf{E}mpirical \textbf{B}ayes estimator for \textbf{MR}, or the MR-EB estimator. 

The hierarchical structure of the parameters is given in Figure \ref{fig4}. The estimation procedure is again via the MCEM algorithm. Implementation details are provided in the Appendix.

\begin{figure}[H]
\centering
\includegraphics[width = 2.5in]{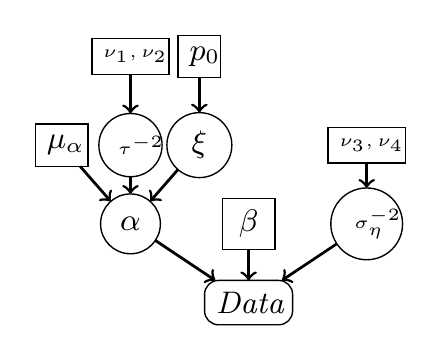}
\caption{The hierarchical structure of parameters under the priors (\ref{bayes1c}), (\ref{bayes1d}), (\ref{prior2a}), and (\ref{prior2b}). All ``squares" represent parameters in the marginal distribution and all ``circles" represent parameters which only appear in the ``complete-data" posterior.
}
\label{fig4}
\end{figure}

\section{Implementation with summary data}
Many public GWAS datasets are available only up to summary statistics for the association studies between individual genetic variants and traits. Moreover, in many cases the data on the interested exposure and that on the interested outcome are available in independent samples. Developing valid methodology for this type of data can broaden the applicability of MR and is of great relevance.

Specifically,  for $j=1,\dots, J$, let $\tilde{\Omega}_j\in \R$ be the association estimate between the interested outcome $Y$ and the $j$-th genetic variant $Z_j$, $\tilde{\sigma}^2_{\Omega,j}\in \R$ be the estimated variance of $\tilde{\Omega}_j$, and $\tilde{\gam}_j \in \R$ be the association estimate between the interested exposure $D$ and $Z_j$. Let $\tilde{\gam}^{(2)}$ be a version of $\tilde{\gam}$ obtained from an independent sample $(Z^{(2)}, D^{(2)})$. We generalize our methods to the case where only $(\tilde{\gam}^{(2)},\tilde{\Omega},\tilde{\sigma}^2_{\Omega})$ are available under some conditions.

Regarding our notation,
\[
   \tilde{\gam}_j = \frac{Z_j^TD}{Z_j^TZ_j} ~~\text{and}~~ \tilde{\Omega}_j = \frac{Z_j^TY}{Z_j^TZ_j},~j=1,\dots J.
\]
 Assuming that $Z^TZ$ is a diagonal matrix, it is easy to see that $\hat{D} = P_ZD = Z \tilde{\gam}$, $P_ZY = Z\tilde{\Omega}$, and $\Var(\tilde{\Omega}_j)= \sig^2_{\eta}/(Z_j^TZ_j)$.
 Thus, the sample moments used throughout the computation can be equivalently represented by the summary statistics, i.e.
 \begin{align}
Z^TZ/\sig^2_{\eta}&= \Sig^{-1}_{\Omega}\label{eq4-1}\\
Z^T\hat{D} /\sig^2_{\eta}&=   \Sig^{-1}_{\Omega}\tilde{\gam}\label{eq4-2}\\
Z^TY/\sig^2_{\eta} &= \Sig^{-1}_{\Omega}\hat{\Omega}\label{eq4-3}\\
\hat{D}^T\hat{D}/\sig^2_{\eta} &=\tilde{\gam}^T\Sig^{-1}_{\Omega}\tilde{\gam}\label{eq4-4}\\
\hat{D}^TY/\sig^2_{\eta} &= \tilde{\gam}^T\Sig^{-1}_{\Omega}\tilde{\Omega},\label{eq4-5}
 \end{align}
 where $\Sig_{\Omega}$ is a diagonal matrix with $(\Sig_{\Omega})_{j,j} = \Var(\tilde{\Omega}_j)$. In our estimation procedure, the unobserved quantities appeared on the left-hand sides of (\ref{eq4-1}) - (\ref{eq4-5}) are replaced by the observed versions of the right-hand sides of (\ref{eq4-1}) - (\ref{eq4-5}), where $\tilde{\gam}$ is replaced by $\tilde{\gam}^{(2)}$ and $\Sigma_{\Omega}$ is replaced by a diagonal matrix with the $j$-th diagonal element equal to $\hat{\sig}^2_{\Omega,j}$. Since the term $\sig^2_{\eta}$ is absorbed into the observed statistics already, it is unnecessary to be updated through the estimation. Thus, we are able to get an MR-EB estimator with summary statistics based on the hierarchical priors (\ref{bayes1c}), (\ref{prior2a}), and (\ref{prior2b}) as in Section \ref{sec: SG} and \ref{sec: MREB}.
\section{Simulations and real studies}
\subsection{Synthetic data experiments}
We evaluate the performance of the proposed methods in comparison to the TSLS and the Lasso estimators in various simulation settings. In particular, we focus on the behavior of MR-EB estimator which is most general and can explore the possibly sparse structure of $\alpha$. The TSLS estimator is computed as a benchmark from classical instrumental variable literatures. The Lasso estimator, which is essentially the sisVIVE estimator in \citep{Lasso},  is proposed to deal with sparse $\alpha$ and hence is added in comparison. The threshold parameter is chosen by 10-fold cross validation as suggested in the paper.

In all the experiments presented in this section, each sample consists of $n = 1000$ observations and $J = 30$ candidate genetic variants.  The genetic variants $Z_i$, $i=1,\dots, n$, are drawn from a multivariate normal distribution with mean zero and identity covariance matrix. The phenotypes $(D_i,Y_i)$,  $i=1,\dots, n$, are generated according to model (\ref{eq1-1}), where each $(v_i,\eps_i)$ is generated from a bivariate normal distribution with mean zero and covariance matrix $\begin{pmatrix} 1 & 0.2 \\ 0.2 &1 \end{pmatrix}$. 

With and without the InSIDE assumption, we allow the following parameters to vary: the strength of causal effect, the distribution of pleiotropic effects and the proportion of invalid instruments. Specifically, we consider two levels of signal strength $\beta\in \{0,0.2\}$, three levels of the mean of pleiotropic effects $\mu_{\alpha}\in \{-0.2, 0, 0.2\}$, and eleven levels of sparsity $p_0\in\{0, 0.1,\dots, 1\}$. In each of these settings, we generate $\gam_j$ from $\rU[0.1, 0.3]$,  $\xi_j$ from $\text{Ber}(1,p_0)$, and $u_j$ from $\rU[\mu_{\alpha}-0.2, \mu_{\alpha} + 0.2]$ in an \textit{i.i.d.} fashion for $j=1,\dots, J$. The pleiotropic effects $\alpha_j=\xi_ju_j$ if the InSIDE assumption is satisfied and $\alpha_j = (0.2\gam_j + u_j)\xi_j$ if the InSIDE assumption is not satisfied, for $j=1,\dots, J$. In each setting, the experiment is independently replicated for 100 times and the mean square error (MSE) is reported. 

As explained before, we take $\nu_0-\nu_4$ to be small numbers. Specifically, we set $\nu_0 = 0.001$, $\nu_1 = 2$, $\nu_2 = 0.4$ and $\nu_3 = \nu_4 = 0.0001$. For the initial values, take $\hat{\beta}^{(0)}=\hat{\mu}^{(0)}_{\alpha}=0$ and $\hat{p}_0^{(0)} = 0.5$.
\begin{figure}[H]
\centering
\includegraphics[height = 3.8cm, width = 11cm]{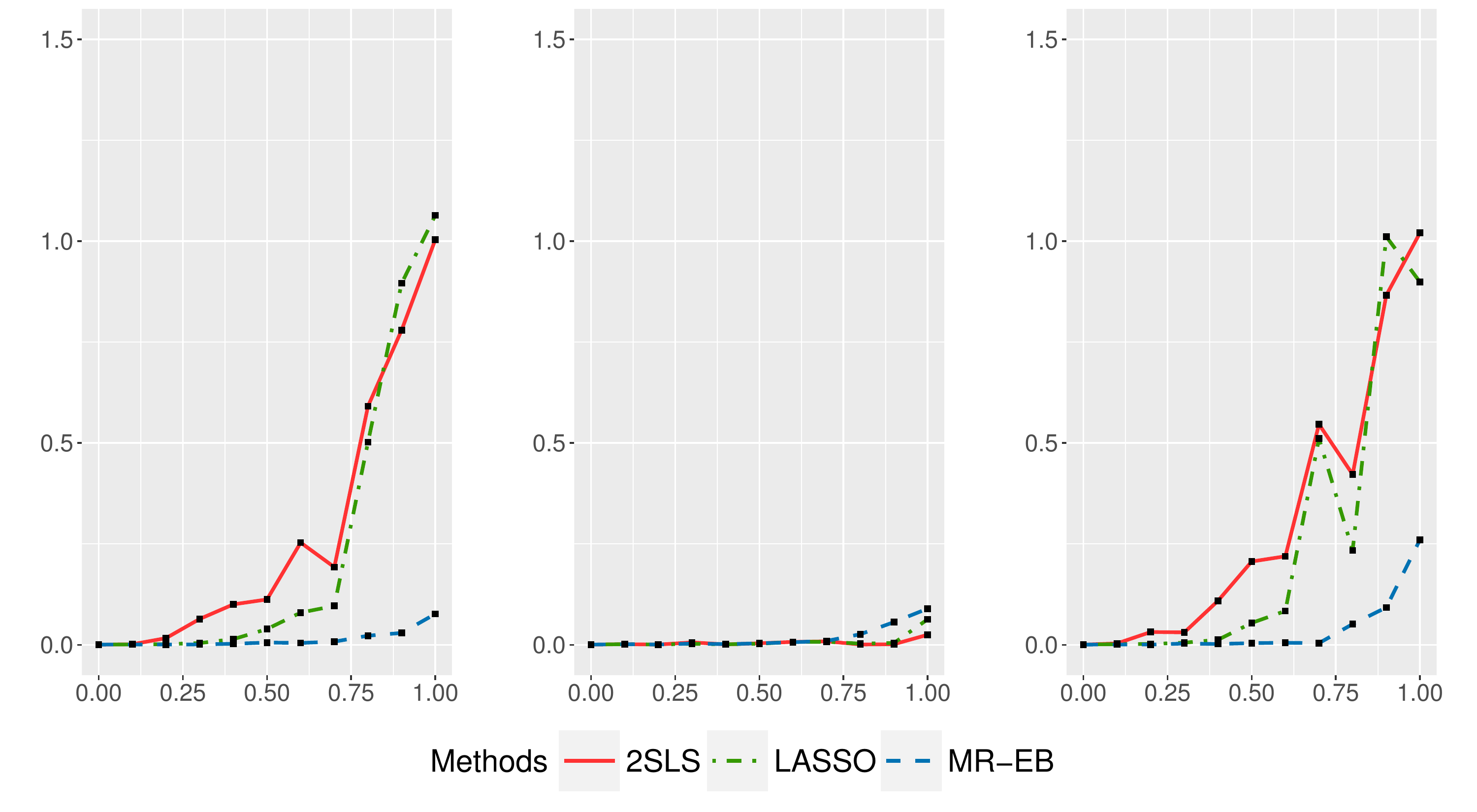}
\fbox{
\includegraphics[height = 3.8cm, width = 3.4cm]{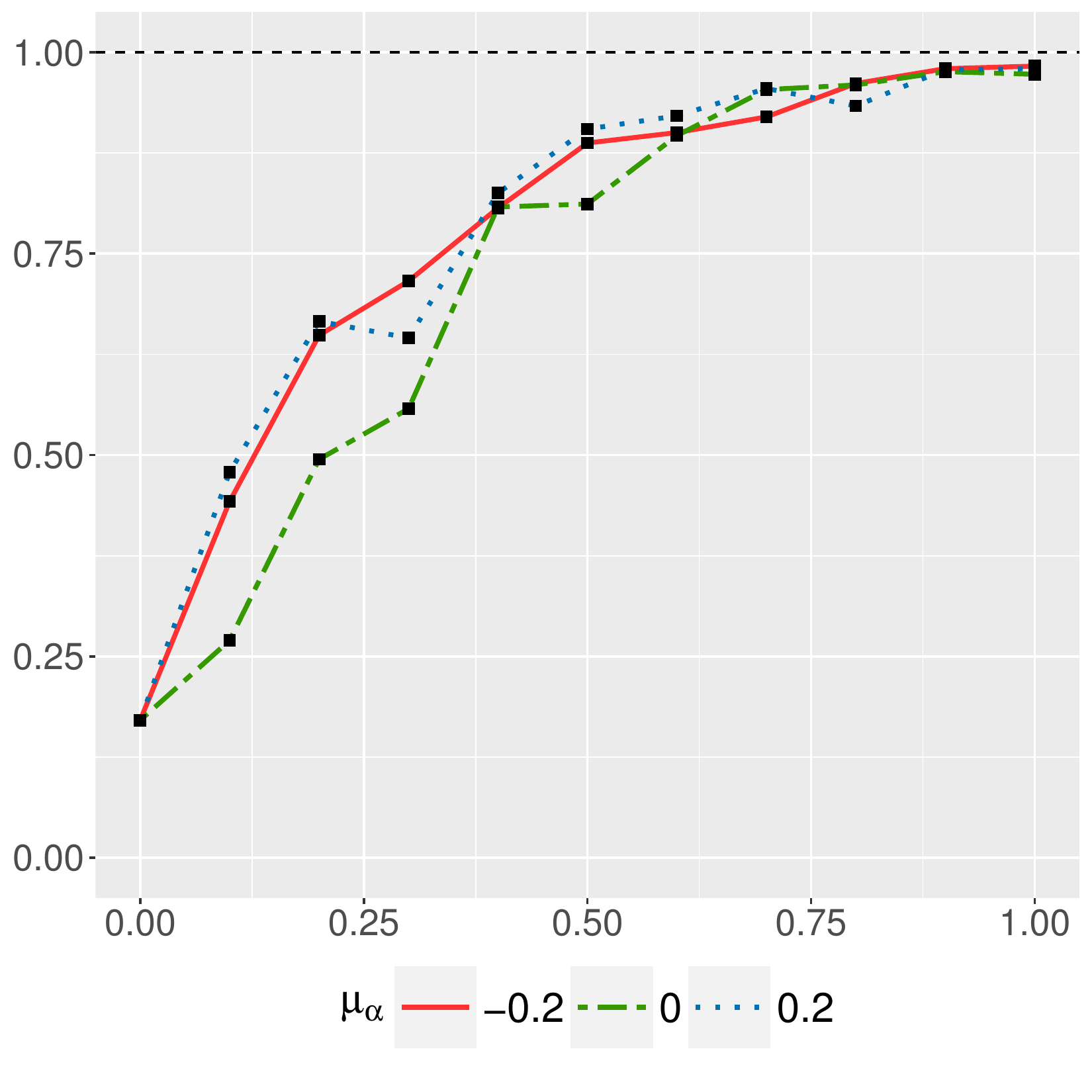}
}
\caption{$ \beta = 0$ and InSIDE assumption is satisfied. The x-axis is the true sparsity level $p_0$ for all the plots.
For the left three plots, each point represents the MSE of 100 experiments for $\mu_{\alpha} = -0.2, 0$, and $0.2$ from left to right, respectively. In the rightmost plot, each point represents the realized $c^{**}$ for the MR-EB estimator in the experiments presented in the left three plots.}
\end{figure}
\begin{figure}[H]
\centering
\includegraphics[height = 3.8cm, width = 11cm]{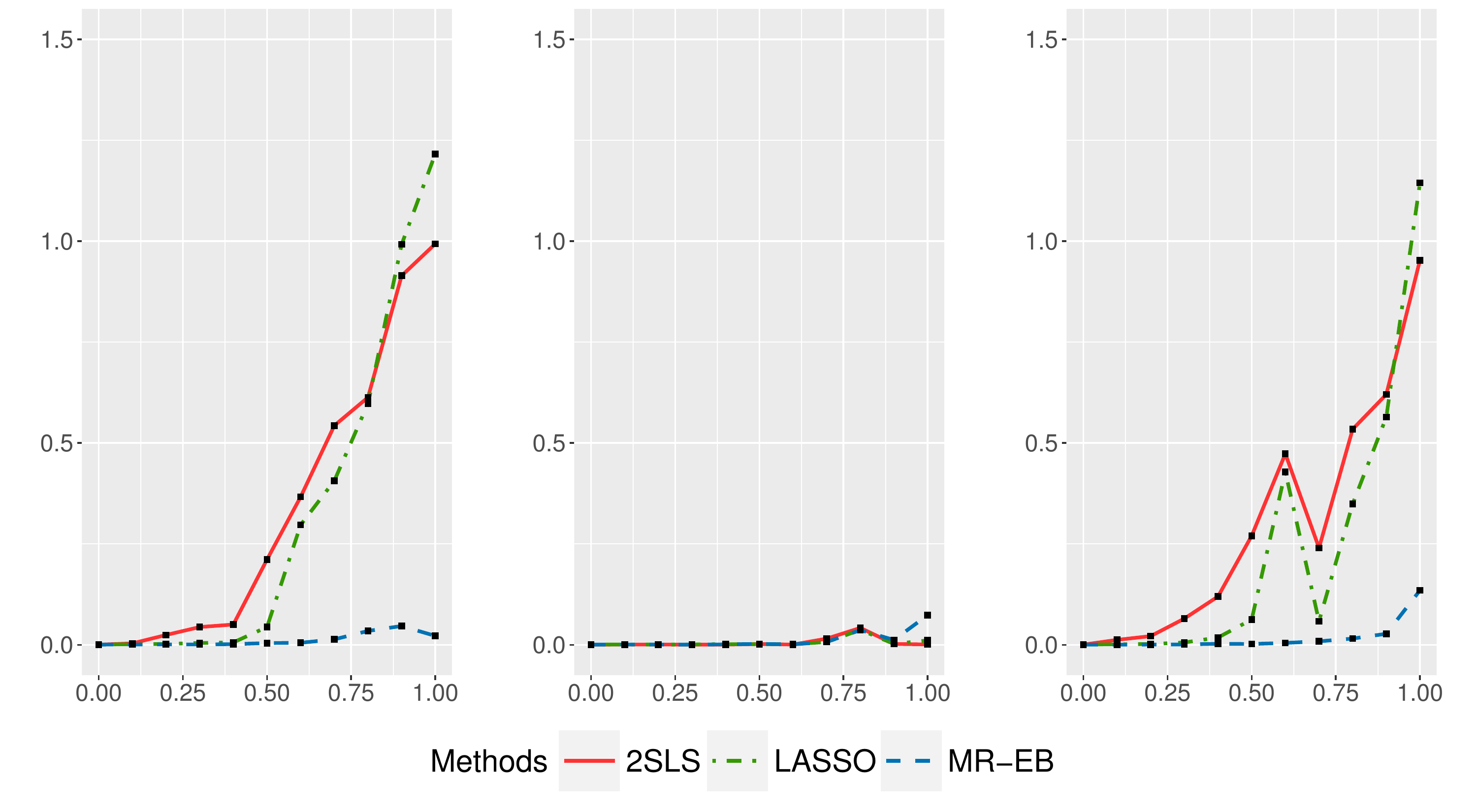}
\fbox{
\includegraphics[height = 3.8cm, width = 3.4cm]{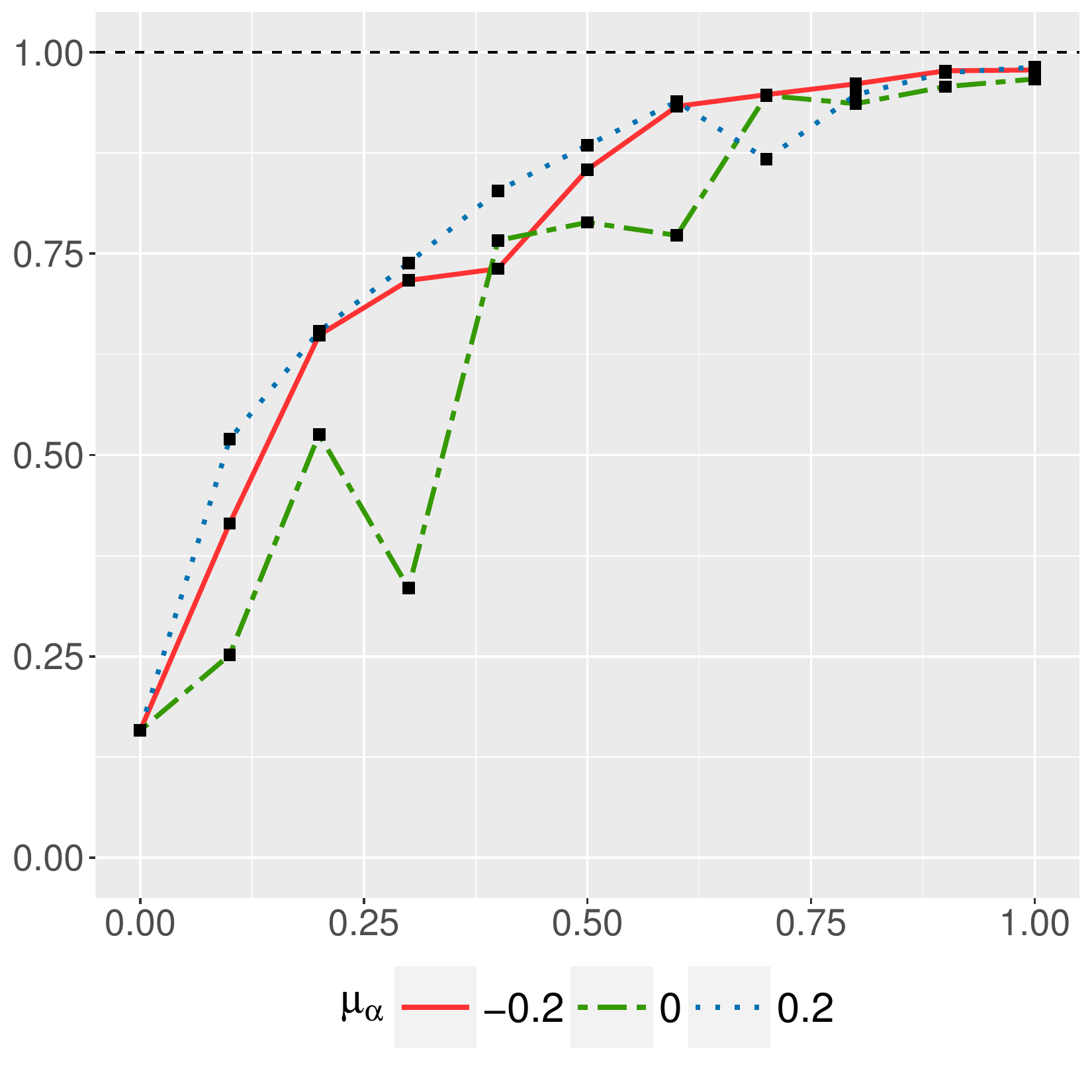}
}
\caption{$ \beta = 0.2$ and InSIDE assumption is satisfied. The x-axis is the true sparsity level $p_0$ for all the plots.
For the left three plots, each point represents the MSE of 100 experiments for $\mu_{\alpha} = -0.2, 0$, and $0.2$ from left to right, respectively.  In the rightmost plot, each point represents the realized $c^{**}$ for the MR-EB estimator in the experiments presented in the left three plots.}
\end{figure}
\begin{figure}[H]
\centering
\includegraphics[height = 3.8cm, width = 11cm]{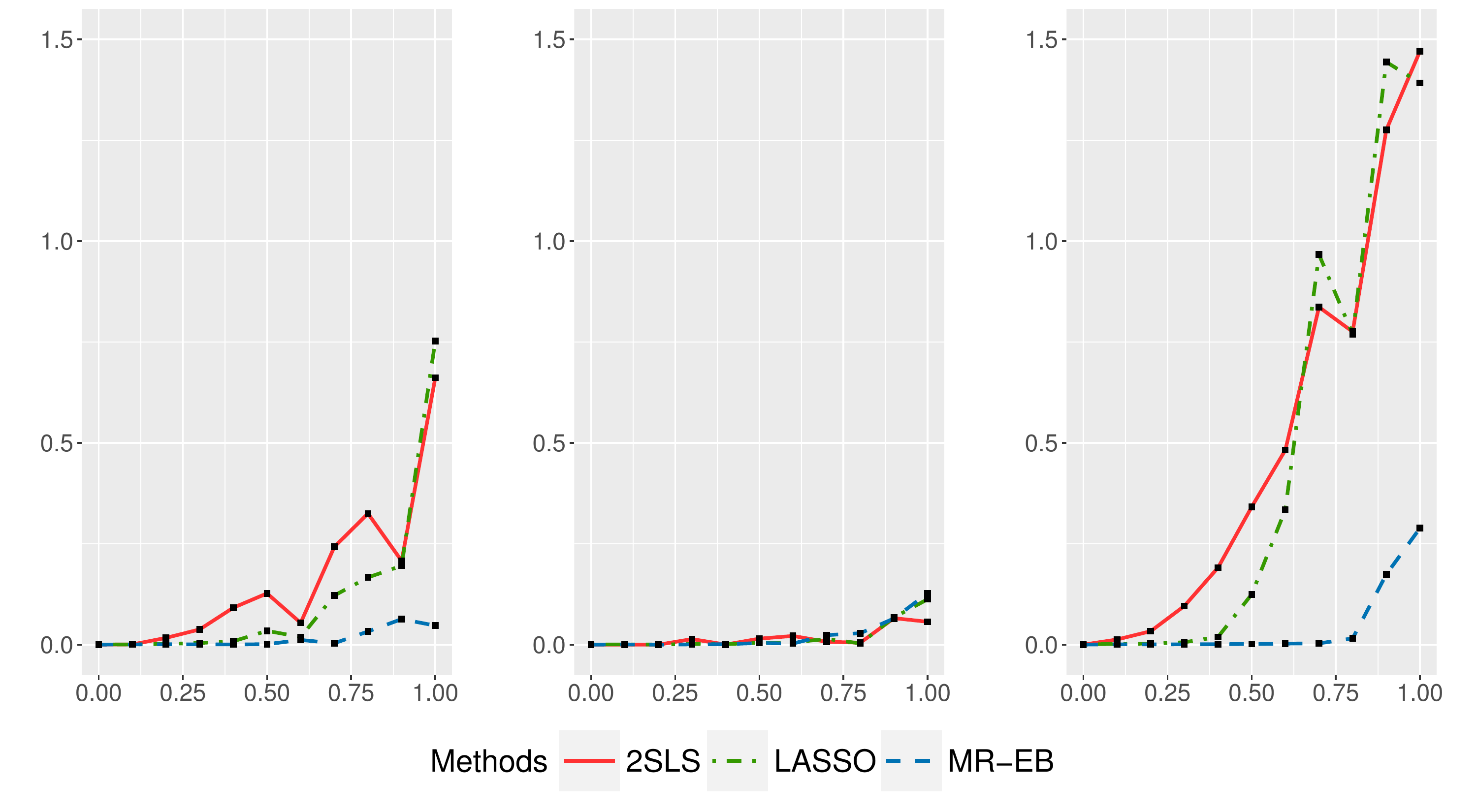}
\fbox{
\includegraphics[height = 3.8cm, width =3.4cm]{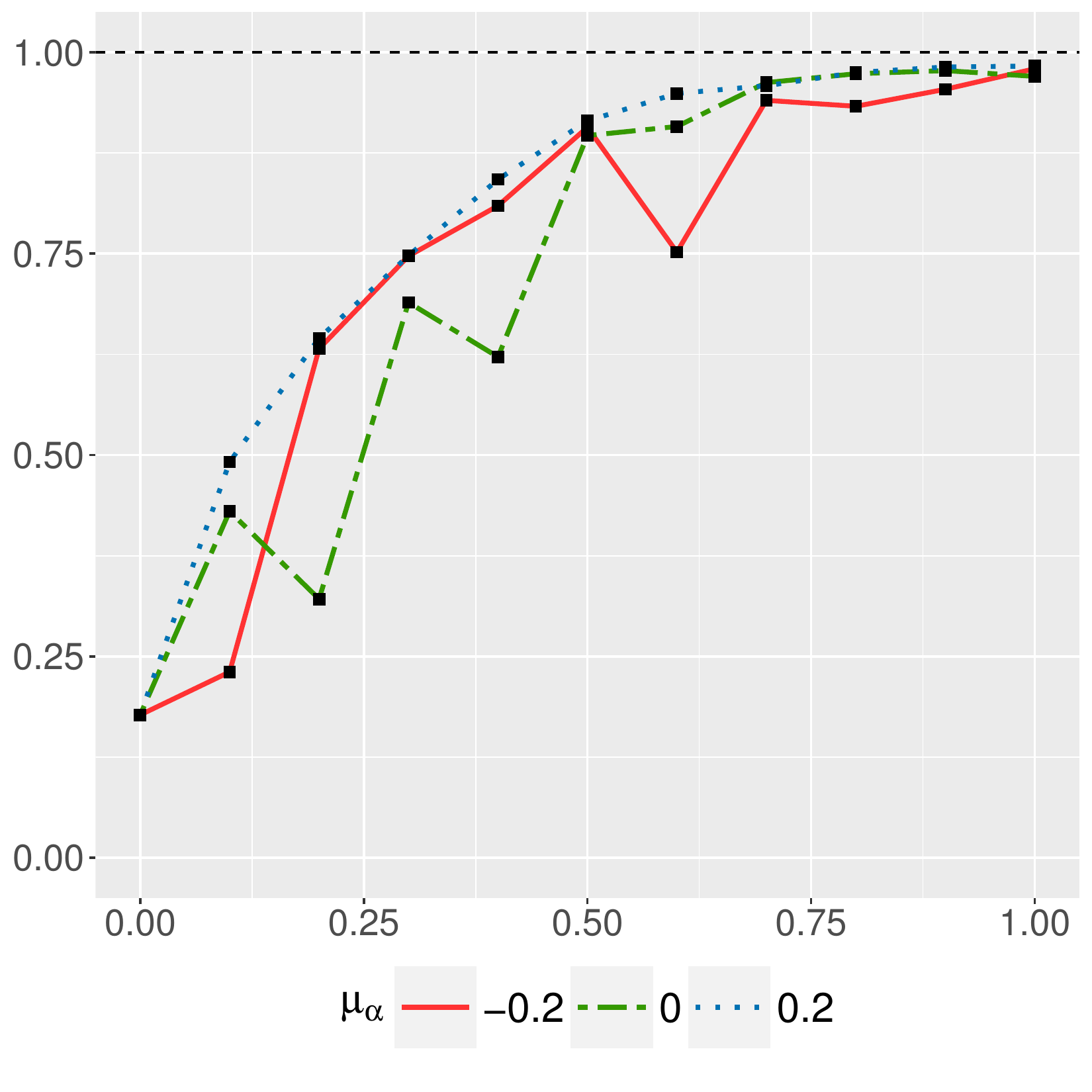}
}
\caption{$ \beta = 0$ and InSIDE assumption is not satisfied. The x-axis is the true sparsity level $p_0$ for all the plots.
For the left three plots, each point represents the MSE of 100 experiments for $\mu_{\alpha} = -0.2, 0$, and $0.2$ from left to right, respectively.  In the rightmost plot, each point represents the realized $c^{**}$ for the MR-EB estimator in the experiments presented in the left three plots.}
\end{figure}
\begin{figure}[H]
\centering
\includegraphics[height = 3.8cm, width = 11cm]{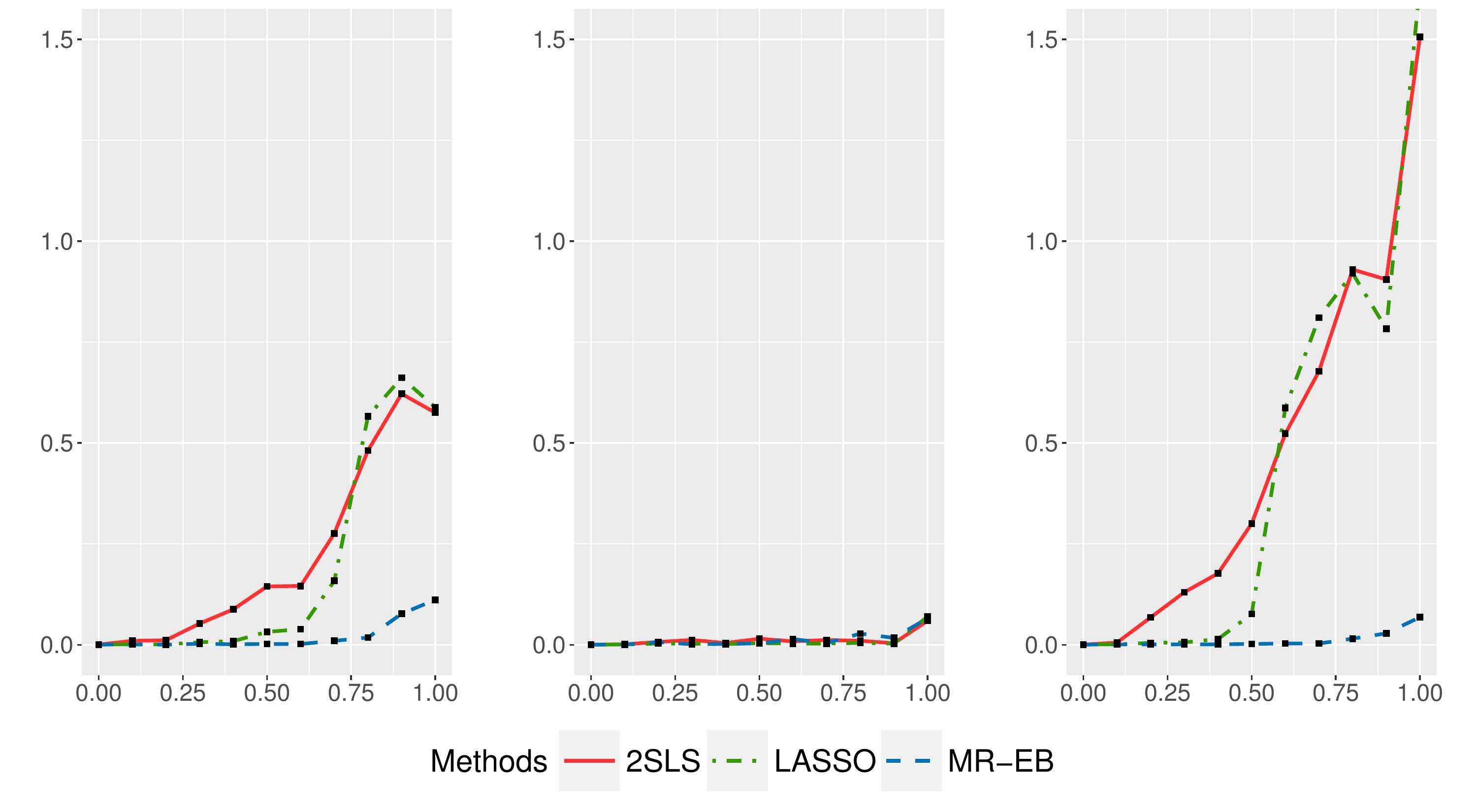}
\fbox{
\includegraphics[height = 3.8cm, width = 3.4cm]{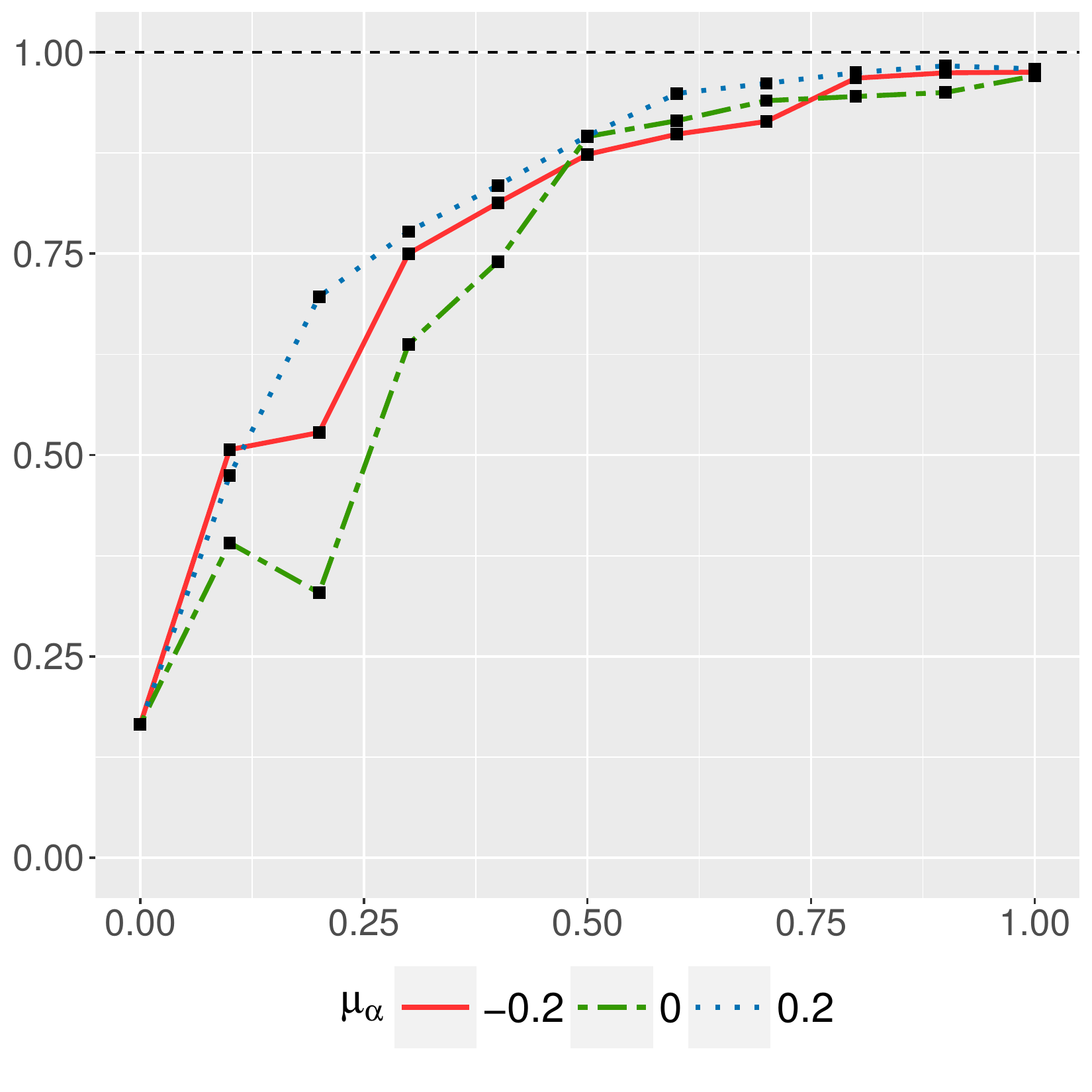}
}
\caption{$ \beta = 0.2$ and InSIDE assumption is not satisfied. The x-axis is the true sparsity level $p_0$ for all the plots.
For the left three plots, each point represents the MSE of 100 experiments for $\mu_{\alpha} = -0.2, 0$, and $0.2$ from left to right, respectively.  In the rightmost plot, each point represents the realized $c^{**}$ for the MR-EB estimator in the experiments presented in the left three plots.}
\end{figure}
From the left three plots in each row, one can see that the MR-EB estimator remains stable in the presence of unbalanced pleiotropic effects, while the other two approaches result in much larger estimation errors. In the balanced pleiotropic effect scenario, the performances of all three methods are comparable. 

Moreover, the number of nonzero pleiotropic effects also plays an important role in the estimation accuracy. One can see that the MR-EB estimator has remarkably reliable performance under different levels of sparseness, even when all the genetic variants are invalid. Though it has relatively large errors when more than $80\%$ of genetic variants are invalid, it is still much more accurate than its counterparts. When the proportion of invalid instruments is less than $50\%$, the proposed estimator is no worse than the Lasso estimator, which shows its stability and adaptivity to the sparsity.
On the other hand, when the proportion of invalid instruments is greater than $50\%$, the Lasso estimator has similar performance as the TSLS estimator and both estimators have large estimation errors. 

Furthermore, the MR-EB estimator has MSE close to zero no matter whether the InSIDE assumption is satisfied or not. In comparison, the MSE of the TSLS estimator and the Lasso estimator change significantly when InSIDE assumption does not hold and the pleiotropic effects are unbalanced. 

The plots of realized $c^{**}$ are to facilitate understanding the performance of the MR-EB estimator. One can see that as the number of nonzero pleiotropic effects increases, $c^{**}$ increases towards 1 superlinearly and it becomes close to 1 when more than $80\%$ of genetic variants are invalid.
\subsection{Case study (i): HDL and type 2 diabetes}
The high density lipoprotein (HDL) cholesterol has the reputation as a "good" cholesterol, since it is negatively associated  in observational studies with the risk of many diseases, for example, myocardial infarction and type 2 diabetes. However, the supporting studies have been unable to control various potential confounders, while the negative association with HDL has lacked convincing biological mechanisms. Hence, the association does not necessarily imply a causal effect.
 
\citet{HDL1} use the traditional MR method to estimate the causality between HDL and the risk of type 2 diabetes. Their results suggest that there is no causal effect of HDL on type 2 diabetes. We access a different set of summary data from MRbase \citep{MRbase} and arrive at a similar conclusion for a related trait. The MRbase is a database and an analytical platform for MR studies, which provides summary data of many published GWAS and some basic analytic tools.

The exposure data is measured plasma HDL cholesterol (unit: mg/dL) from the Global Lipids Genetics Consortium \citep{WilerCJ13} with a sample size 187167. The outcome data is measured fasting glucose (unit: mmol/L) from the Meta-Analyses of Glucose and Insulin-related traits Consortium \citep{Dupuis10} with a sample size 46186. Hyperglycemia in the fasting state is one of the criteria that defines type 2 diabetes \citep{Diabete2}. Thus, fasting glucose is an important indicator of Type 2 diabetes.

For the analysis, 83 genetic variants are selected and harmonized automatically by the MRbase, which excludes linkage disequilibrium and selects variants which are robustly associated with the target traits with the genome-wide significance threshold $5\times 10^{-8}$.

Four method are applied on this dataset. The estimate given by TSLS is -0.0282 mmol/L per mg/dL; the estimate given by the Egger's regression \citep{Egger}  is -0.0345 mmol/L per mg/dL; the estimate given by the inverse-variance weighted median estimator \citep{IVWM} is -0.0290 mmol/L per mg/dL; the estimate given by the MR-EB estimator is -0.0312 mmol/L per mg/dL, where the hyper-parameters $\nu_0-\nu_2$ are specified as 0.001, 2, and 0.2, respectively. One can see that these methods generate similar estimates for this dataset.

\subsection{Case study (ii): LDL and type 2 diabetes}
As introduced at the beginning of the paper, we study the causal effect of LDL cholesterol on the type 2 diabetes in this section.
 
The exposure data is measured plasma LDL cholesterol (unit: mg/dL) from the Global Lipids Genetics Consortium \citep{WilerCJ13}  with a sample size 173082. The outcome data is the fasting glucose (unit: mmol/dL) which is from the same source of data as in case study (i). For the analysis, 72 genetic variants are selected and harmonized.

For this dataset, the estimate given by TSLS is -0.0157 mmol/L per mg/dL; the estimate given by the Egger's regression \citep{Egger} is -0.0248 mmol/L per mg/dL; the estimate given by the inverse-variance weighted median estimator \citep{IVWM} is -0.0121 mmol/L per mg/dL; the estimate given by the MR-EB estimator is -0.0038 mmol/L per mg/dL, for which the hyper-parameters $\nu_0-\nu_2$ are specified as 0.001, 2, and 0.2, respectively. 
\section{Discussion}
In this paper, we have developed empirical Bayes hierarchical models for estimating the causal effect in the presence of invalid instruments for MR studies. Due to the confoundness of pleiotropic effects, the causal effect cannot be identified with the traditional TSLS estimator. Instead of making structural assumptions about the unobserved pleiotropic effects (such as the ``partially invalid" assumption and the InSIDE assumption), we set up hierarchical models, which utilize the regularization structure to share information across the hierarchy. Empirical Bayes approaches are employed to estimate the unknown marginal parameters, which take care of the unbalanced pleiotropic effects and the unknown sparsity. Theoretically, we have developed empirical error bounds which shed light on a class of shrinkage estimators. 
The resulting MR-EB estimator can be efficiently implemented with the MCEM algorithm. 

The simulation results demonstrate the reliable and compelling performance of the MR-EB estimator throughout different levels of signal strength and sparsity, in comparison to existing methods. The proposed method remains reliable in the presence of unbalance pleiotropic effects and when the InSIDE assumption is not satisfied. The simulation results are consistent with the theorems and discussions presented in the previous section.

There are still interesting and open problems in the scope of current topic. Firstly, many epidemiological studies are interested in the causal effect of exposures on the risk of certain diseases. Thus, it is an important task to provide a reliable procedure to estimate the causal effect for the binary outcome data, or equivalently the probability of occurrence of an event. Under some model assumptions, one can generalize the proposed approach to the logistic and probit models. Secondly, this paper together with many previous works have been focusing on the estimation procedure, while generating valid interval estimates with mild conditions and cheap computation remains to be a challenging and worthwhile topic. Thirdly, to further reduce the assumptions on the prior distribution, one may also consider fitting a nonparametric empirical Bayes model.  

The R code of the implementation of the proposed methods is available from the author upon request.

\section*{Acknowledgement}
The author gratefully thanks Steven Buyske for insightful discussions and valuable comments on both the methodology and the presentation.


%



\appendix
\section*{Appendix}
\section{Proof of theorems and lemmas}
\label{sec: Ap-A}
\subsection{Proof of Theorem \ref{thm1}}
\label{sec: Ap-A1}
\begin{proof}
Define $\Gam = \tau^2I_{J\times J}$ and $\Xi = \begin{pmatrix}
\hat{D}^T\hat{D}/\sig^2_{\eta} &\hat{D}^TZ/\sig^2_{\eta}\\
Z^T\hat{D}/\sig^2_{\eta} & Z^TZ/\sig^2_{\eta}+\Gam^{-1}
\end{pmatrix}^{-1}$.
From (\ref{func1}), we can get
\begin{align*}
\begin{pmatrix}
\hat{\beta}^{\mu_{\alpha}}  \\ \hat{\alpha}^{\mu_{\alpha}}
\end{pmatrix}&= \Xi
\begin{pmatrix}
\hat{D}^TY/\sig^2_{\eta}\\
\hat{Z}^TY/\sig^2_{\eta}+ \Gam^{-1}\mu_{\alpha}
\end{pmatrix}\\
&= \Xi
\begin{pmatrix}
 \hat{D}^TY/\sig^2_{\eta}\\
Z^TY/\sig^2_{\eta}  + \Gam^{-1}\alpha + \Gam^{-1}(\mu_{\alpha}-\alpha)
\end{pmatrix}\\
&= \begin{pmatrix}
\beta\\
\alpha
\end{pmatrix} + \Xi\begin{pmatrix}
\hat{D}^T\hat{\eta}/\sig^2_{\eta}\\
 Z^T\hat{\eta}/\sig^2_{\eta} + \Gam^{-1}(\mu_{\alpha}-\alpha)
\end{pmatrix}.
\end{align*} 

 From the above derivation, we can obtain that
\begin{align}
\hat{\beta}^{\mu_{\alpha}}  &=\beta + \Xi_{1,1}\hat{D}^T\hat{\eta}/\sig^2_{\eta} + \Xi_{1,J}(Z^T\hat{\eta}/\sig^2_{\eta}+\Gam^{-1}\left(\mu_{\alpha}-\alpha\right))\nonumber\\
&=\beta + \underbrace{\Xi_{1,1}\hat{D}^T\hat{\eta}/\sig^2_{\eta} +\Xi_{1,J}Z^TP_{\hat{D}}\hat{\eta}/\sig^2_{\eta}}_{E_1}+ \underbrace{\Xi_{1,J}(Z^T\PhDperp\hat{\eta}/\sig^2_{\eta}+\Gam^{-1}\left(\mu_{\alpha}-\alpha\right))}_{E_2}\label{thm1-1}.
\end{align}
By the matrix inverse formula and some simple algebra, we can get
\begin{align}
\Xi_{J,J} &=  \left(Z^TZ/\sig^2_{\eta}+\Gam^{-1} - \frac{Z^T\hat{D}\hat{D}^TZ/\sig^2_{\eta}}{\hat{D}^T\hat{D}}\right)^{-1}= (B-A)^{-1}\label{thm1-2}\\
\Xi_{1,J} &=-\frac{\hat{D}^TZ\Xi_{J,J}}{\hat{D}^T\hat{D}} = -\frac{\sig^2_{\eta}\hat{\gam}^TA(B-A)^{-1}}{\hat{D}^T\hat{D}},\label{thm1-3}
\end{align}
where the last step is due to $\hat{D}^TZ = \hat{D}^TP_{\hat{D}}Z=\hat{\gam}^TZ^TP_{\hat{D}}Z$ and (\ref{thm1-2}).

Thus, for $E_1$ in (\ref{thm1-1}), we have
\begin{align}
 E_1& = \Xi_{1,1}\hat{D}^T\hat{\eta}/\sig^2_{\eta} +\frac{\Xi_{1,J}Z^T\hat{D}\hat{D}^T\hat{\eta}/\sig^2_{\eta}}{\hat{D}^T\hat{D}}\nonumber\\
   &=\left(\Xi_{1,1} -\Xi_{1,J}\Xi^{-1}_{J,J}\Xi_{J,1}\right)\hat{D}^T\hat{\eta}/\sig^2_{\eta}\nonumber\\
   &= \frac{\hat{D}^T\hat{\eta}}{\hat{D}^T\hat{D}}\label{thm1-4},
   \end{align}
   where the second equality can be seen from the first part of (\ref{thm1-3}) and the third equality is again by the matrix inverse formula.
 \begin{align}
|E_2|&=\left|-\frac{\sig^2_{\eta}\hat{\gam}^TA(B-A)^{-1}(Z^T\PhDperp\hat{\eta}/\sig^2_{\eta}+\Gam^{-1}\left(\mu_{\alpha}-\alpha\right))}{\hat{D}^T\hat{D}}\right|\nonumber\\
&\leq\frac{\Lambda_{max}(A(B-A)^{-1})\left(\sig^2_{\eta}\|\hat{\gam}\|_2\|\Gam^{-1}(\alpha-\mu_{\alpha})\|_2+\|\hat{\gam}\|_2\|Z^T\PhDperp \hat{\eta}\|_2\right)}{\hat{D}^T\hat{D}}\nonumber\\
&\leq\frac{\sig^2_{\eta}c^*\|\hat{\gam}\|_2\|\Gam^{-1}(\alpha-\mu_{\alpha})\|_2}{(1-c^*)\hat{D}^T\hat{D}}+ \frac{c^*\|\hat{\gam}\|_2\|Z^T\PhDperp \hat{\eta}\|_2}{(1-c^*)\hat{D}^T\hat{D}} \nonumber\\
&=\frac{\sig^2_{\eta}c^*\|\hat{\gam}\|_2\|\alpha-\mu_{\alpha}\|_2}{\tau^2(1-c^*)\hat{D}^T\hat{D}}+ \frac{c^*\|\hat{\gam}\|_2\|Z^T\PhDperp \hat{\eta}\|_2}{(1-c^*)\hat{D}^T\hat{D}} \label{thm1-5},
\end{align}
where the last inequality is due to $A(B-A)^{-1} = AB^{-1}(I- AB^{-1})^{-1}$, $\Lambda_{max}(AB^{-1}(I- AB^{-1})^{-1})\leq \frac{c^*}{1-c^*}<\infty$, and $0<c^* <1$.

Thus, by (\ref{thm1-1}), (\ref{thm1-4}), and (\ref{thm1-5}), we have
\begin{align*}
  |\hat{\beta}^{\mu_{\alpha}} - \beta|&\leq |E_1| + |E_2|\\
  &\leq \frac{\sig^2_{\eta}c^*\|\hat{\gam}\|_2\|\alpha-\mu_{\alpha}\|_2}{\tau^2(1-c^*)\hat{D}^T\hat{D}}+ \frac{c^*\|\hat{\gam}\|_2\|Z^T\PhDperp \hat{\eta}\|_2}{(1-c^*)\hat{D}^T\hat{D}} + |\frac{\hat{D}^T\hat{\eta}}{\hat{D}^T\hat{D}}|.
\end{align*}
\end{proof}

\subsection{Proof of Lemma \ref{lem1}}
\label{sec: Ap-A2}
\begin{proof}
The matrices $A$ and $B$ satisfy that
\[
    AB^{-1} = I - \frac{1}{n}(Z^T\PhDperp Z  + \frac{\sig^2_{\eta}}{\tau^2}I_{J\times J})B^{-1},
\]
where $\frac{1}{n}(Z^T\PhDperp Z  + \frac{\sig^2_{\eta}}{\tau^2}I_{J\times J})$ and $B$ are both positive definite matrices for $0<\sig^2_{\eta}/\tau^2<\infty$.
Thus, 
\[
   \Lambda_{\max}(AB^{-1} ) = 1 - \Lambda_{\min}(\frac{1}{n}(Z^T\PhDperp Z  + \frac{\sig^2_{\eta}}{\tau^2}I_{J\times J})B^{-1})<1.
\]

One the other hand, matrix $A$ is a  rank-1 semi-definite matrix and
\begin{align*}
\Lambda_{\max}(A)&= \text{Trace}(A) \\
&= \frac{1}{n}\frac{\hat{D}^TZ^TZ\hat{D}}{\hat{D}^T\hat{D}}>0,
\end{align*}
if $\|\hat{D}\|_2>0$ and $\Lambda_{\min} (Z^TZ/n) > 0$. 

Let $u_A$ be such that $u_A = \argmax_{\|u\|_2 = 1} u^TAu$. By the well-known variational theorem, we have
\begin{align*}
 \Lambda_{\max}(AB^{-1})&= \Lambda_{\max}(B^{-1/2}AB^{-1/2}) \\
 &=\max_{\|u\|_2=1} u^TB^{-1/2}AB^{-1/2}u \\
 &\geq u_A^TAu_A/u_A^TBu_A\\
  & \geq \Lambda_{\max}(A)\Lambda^{-1}_{\max}(B),
\end{align*}
where the third step is by taking $u = B^{1/2}u_A/(u_A^TBu_A)^{1/2}$ and $u_A^TBu_A>0$ due to $B$ is a positive definite matrix.
\end{proof}

\subsection{Proof of Corollary \ref{cor1}}
\label{sec: Ap-A3}
\begin{proof}
The proof follows the line of the proof of Theorem \ref{thm1} and is omitted here.
\end{proof}

\section{Implementation details}
\label{sec: Ap-B}
\subsection{Single Gaussian prior}
\label{sec: Ap-B1}
In this section, we discuss the estimation procedure under the hierarchical model (\ref{bayes1a}) - (\ref{bayes1d}). 

Start with initial values $(\hat{\beta}^{(0)},\hat{ \mu}^{(0)}_{\alpha})$. 

\textbf{E-step}:
At round t, generate $(\alpha^{(t)}_i,(\tau^2)^{(t)}_i, (\sig^2_{\eta})^{(t)}_i), i = 1\dots ,m$, from $p(\alpha, \tau^2, \sig^2_{\eta}| \mathcal{D},\hat{\beta}^{(t-1)},\hat{\mu}_{\alpha}^{(t-1)})$ by the Gibbs sampling procedure:
\begin{align}
\alpha^{(t)}_i &\sim p(\alpha|\mathcal{D}, \hat{\beta}^{(t-1)},\hat{\mu}_{\alpha}^{(t-1)},(\sig_{\eta}^2)^{(t)}_{i-1}, (\tau^2)^{(t)}_{i-1})\label{ap1-1a}\\
(\tau^2)^{(t)}_i &\sim p(\tau^2|\mathcal{D}, \hat{\beta}^{(t-1)},\hat{\mu}_{\alpha}^{(t-1)},\alpha^{(t)}_{i},(\sig_{\eta}^2)^{(t)}_{i-1})\label{ap1-1b}\\
(\sig^2_{\eta})_i^{(t)} &\sim p(\sig^2_{\eta}|\mathcal{D}, \hat{\beta}^{(t-1)},\hat{\mu}_{\alpha}^{(t-1)},\alpha^{(t)}_{i},(\tau^2)^{(t)}_{i-1}) . \label{ap1-1c}
\end{align}
Specifically, the sampling distributions in (\ref{ap1-1a}) -(\ref{ap1-1c}) are
\begin{align*}
\alpha_i^{(t)}&\sim N(\underline{\theta}_\alpha, \underline{\Sigma}_{\alpha}) \\
(\tau^{-2})_i^{(t)}&\sim \text{Gamma}(\nu_1 + \frac{J}{2}, \nu_2 + \frac{\sum_{j=1}^J(\alpha^{(t)}_{i,j}-\hat{\mu}^{(t-1)}_{\alpha})^2}{2}) \\
(\sig^{-2}_{\eta})_i^{(t)} &\sim \text{Gamma}(\nu_3 + \frac{n}{2},\nu_4 +\frac{1}{2}\|Y - \hat{D}\hat{\beta}^{(t-1)} - Z\alpha_i^{(t)}\|^2_2),
\end{align*}
where $ \underline{\Sigma}_{\alpha} =\left( Z^TZ/(\sig^2_{\eta})_{i-1}^{(t)}+  (\Gam^{(t)}_{i-1})^{-1}\right)^{-1}$ and $\underline{\theta}_\alpha = \underline{\Sigma}_{\alpha} (Z^T (Y - \hat{D}\hat{\beta}^{(t-1)})/(\sig^2_{\eta})_{i-1}^{(t)} + (\Gam^{(t)}_{i-1})^{-1}\hat{\mu}^{(t-1)}_{\alpha})$ for $(\Gam^{(t)}_{i-1})= (\tau^2)^{(t)}_{i-1}I_{J\times J}$.

\textbf{M-step}: Compute
\[
   (\hat{\beta}^{(t)}, \hat{\mu}^{(t)}_{\alpha}) = \argmax_{(\beta,\mu_{\alpha})\in \R^{2}}  \frac{1}{m}\sum_{i=1}^m \log p(\beta,\mu_{\alpha}|\mathcal{D},\alpha^{(t)}_i,(\tau^2)^{(t)}_i, (\sig^2_{\eta})^{(t)}_i).
\]
 The maximizers take the form
\begin{align*}
\hat{\beta}^{(t)} &= \frac{1}{m}\sum_{i=1}^m \frac{\hat{D}^T(Y- Z\hat{\alpha}_i^{(t)})}{\hat{D}^T\hat{D}}\\
\hat{\mu}^{(t)}_{\alpha} &=\frac{\sum_{i=1}^m \sum_{j=1}^J \hat{\alpha}_{i,j}^{(t)}/(\tau^{2})_i^{(t)}}{\sum_{i=1}^m J/(\tau^{2})_i^{(t)}}.
\end{align*}

Iteratively operate the E-step and M-step until it converges. At the convergence of $(\hat{\beta}^{(t)},\hat{\mu}^{(t)}_{\alpha})$, produce $\hat{\beta}^{(t)}$ as the final estimate of $\beta$. 

\subsection{Mixture Gaussian prior}
\label{sec: Ap-B2}
In this section, we discuss the estimation of MR-EB estimator under prior  (\ref{bayes1c}), (\ref{bayes1d}), (\ref{prior2a}), and (\ref{prior2b}). We can still apply the MCEM algorithm with some modifications.

Start with initial values $(\hat{\beta}^{(0)}, \hat{\mu}^{(0)}_{\alpha},\hat{p}_0^{(0)})$.

\textbf{E-step}: Generate $(\alpha^{(t)}_i, \xi^{(t)}_i, (\sigma_{\eta}^2)^{(t)}_i,\tau^{(t)}_i), i = 1\dots ,m$, from $p(\alpha, \xi ,\sigma^2_{\eta},\tau^2| \mathcal{D}, \hat{\beta}^{(t-1)},\hat{\mu}^{(t-1)}_{\alpha},\hat{p}^{(t-1)}_0)$ by Gibbs sampling procedure:
\begin{align*}
&\alpha_i^{(t)}\sim N(\underline{\theta}_{\alpha,\xi}, \underline{\Sigma}_{\alpha,\xi}) \\
&\xi_{i,j}^{(t)}\sim Ber(\underline{p_j}), ~j= 1,\dots, p\\
&  (\tau^{-2})_i^{(t)} \sim \text{Gamma}(\nu_1 + \frac{J}{2}, \nu_2 + \frac{1}{2}\sum_{j=1}^J\frac{(\alpha^{(t)}_{i,j}-\hat{\mu}^{(t-1)}_{\alpha}\xi^{(t)}_{i,j})^2 }{ (1-\nu_0)\xi^{(t)}_{i,j}+\nu_0})\\
&  (\sig^{-2}_{\eta})_i^{(t)} \sim \text{Gamma}(\nu_3 + \frac{n}{2},\nu_4 +\frac{1}{2}\|Y - \hat{D}\hat{\beta}^{(t-1)} - Z\alpha_i^{(t)}\|^2_2),
\end{align*}
where $ \underline{\Sigma}_{\alpha,\xi} =\left( Z^TZ/ (\sig^2_{\eta})_{i-1}^{(t)} +  (\Gam^{(t)}_{\xi,i-1})^{-1}\right)^{-1}$, $\underline{\theta}_{\alpha,\xi} = \underline{\Sigma}_{\alpha,\xi} (Z^T(Y-\hat{D}\hat{\beta}^{(t-1)})/ (\sig^2_{\eta})^{(t)}_{i-1} + (\Gam_{\xi,i-1}^{(t)})^{-1}\hat{\mu}^{(t-1)}_{\alpha})$, and
 $\underline{p_j}  = \frac{\hat{p}^{(t-1)}_0\phi(\hat{\alp}^{(t)}_{i,j}|\hat{\mu}^{(t-1)}_{\alp},(\tau^2)^{(t)}_{i-1})}{\hat{p}^{(t-1)}_0\phi(\hat{\alp}^{(t)}_{i,j}|\hat{\mu}^{(t-1)}_{\alp},(\tau^2)^{(t)}_{i-1}) +  (1-\hat{p}^{(t-1)}_0)\phi(\hat{\alp}^{(t)}_{i,j}|0,\nu_0(\tau^2)^{(t)}_{i-1})}$ for a diagonal matrix $\Gam_{\xi,i-1}^{(t)}$ with  $(\Gam_{\xi,i-1}^{(t)})_{j,j}= (\nu_0 + (1-\nu_0)\xi_{i-1}^{(t)})(\tau^2)^{(t)}_{i-1}$.

\textbf{M-step}:
\[ (\hat{\beta}^{(t)},\hat{\mu}^{(t)}_{\alpha},\hat{p}^{(t)}_0)= \argmax_{(\beta,\mu_{\alpha},p_0)} \frac{1}{m}\sum_{i=1}^m \log p(\beta,\mu_{\alpha},p_0|\mathcal{D},\alpha^{(t)}_i, \xi^{(t)}_i,(\tau^2)^{(t)}_i,(\sig^2_{\eta})^{(t)}_i).
\]

 The maximizers in the M-step take the form
\begin{align*}
\hat{\beta}^{(t)} &= \frac{1}{m}\sum_{i=1}^m \frac{\hat{D}^T(Y- Z\hat{\alpha}_i^{(t)})}{\hat{D}^T\hat{D}}\\
\hat{\mu}^{(t)}_{\alpha} &=\frac{\sum_{i=1}^m\sum_{j=1}^J \hat{\alpha}_{i,j}^{(t)}\hat{\xi}_{i,j}^{(t)}/(\tau^2)^{(t)}_i}{\sum_{i=1}^m\sum_{j=1}^J \hat{\xi}_{i,j}^{(t)}/(\tau^2)^{(t)}_i} \\
\hat{p}^{(t)}_0 &= \frac{1}{mJ}\sum_{i=1}^m \sum_{j=1}^J \hat{\xi}^{(t)}_{i,j}.
\end{align*}

At the convergence of $(\hat{\beta}^{(t)},\hat{\mu}^{(t)}_{\alpha},\hat{p}^{(t)})$, produce $\hat{\beta}^{(t)}$ as the final estimate of $\beta$.

\end{document}